\documentstyle[aps,preprint]{revtex}
\begin{document}
\draft

\title{Magnetically charged solutions via an analog of 
       the electric-magnetic 
       duality in (2+1)-dimensional gravity theories}
\author{Youngjai Kiem$^{(a)}$ and Dahl Park$^{(b)}$ }
\address{$^{(a)}$ Department of Physics \\
                  Sejong Univerity\\
                  Seoul 143-747, KOREA \\
                  E-mail: ykiem@phy.sejong.ac.kr \\
         $^{(b)}$ Department of Physics \\
                  KAIST\\
                  Taejon 305-701, KOREA \\
                  E-mail: dpark@chep6.kaist.ac.kr}
\maketitle

\begin{abstract}
We find an analog of the electric-magnetic duality, which is a $Z_2$ 
transformation between magnetic and electric sectors of the static and 
rotationally symmetric solutions in a class of (2+1)-dimensional 
Einstein-Maxwell-Dilaton gravity theories.  The theories in our 
consideration include, in particular, one parameter class of theories 
continuously connecting the Banados-Teitelboim-Zanelli (BTZ) gravity 
and the low energy string effective theory.  When there is no $U(1)$ 
charge, we have $O(2)$ or $O(1,1)$ symmetry, depending on a parameter
that specifies each theory.  Via the $Z_2$ transformation, we obtain 
exact magnetically charged solutions from the known electrically 
charged solutions.  We explain the relationship between the $Z_2$ 
transformation and $O(2,Z)$ symmetry, and comment on the $T$-duality 
of the string theory.
\end{abstract}

\pacs{04.20.Jb, 04.60.Kz, 11.30.Er}

In general, we do not expect the existence of 
the electric-magnetic duality in a 
(2+1)-dimensional theory.  This is based on an observation that,
unlike the (3+1)-dimensional case \cite{horowitz}, the 
number of independent components for the electric field in 
(2+1)-dimension is different from that of the 
magnetic field.  If we restrict 
our attention only to static and rotationally symmetric field 
configurations in (2+1)-dimensional theories, however, we have the same 
number of the electric-magnetic field component and it is conceivable
that some analog of the electric-magnetic duality 
may exist \cite{emd3}.  There are 
some reasons why we are interested in this issue.  The $T$-duality
of the string theory has been increasingly playing a significant role in 
recent developments of the string theory \cite{stringd}.  Since the 
target space effective action of the string theory contains $U(1)$ 
gauge fields from the open string sector, the 
$T$-duality may imply some analog of the 
electric-magnetic duality in a target space geometry.  Conversely,
the study of the electric-magnetic duality may lead to a better
understanding of the $T$-duality in the low energy string theory.  
In particular, 
Cadoni \cite{caldoni} recently found an $O(2)$ symmetry in a class 
of (2+1)-dimensional Kaluza-Klein type theories that includes the low 
energy string effective theory without $U(1)$ gauge fields and 
the (uncharged) Banados-Teitelboim-Zanelli (BTZ) theory \cite{btz}
\cite{carlip}. It was further suggested there that the discrete 
version of the 
symmetry, i.e., $O(2,Z)$, may be related to the $O(2,2,Z)$ duality 
from the string theory description of the (2+1)-dimensional
gravity theories \cite{stdual}.  There is also a more practical reason 
why we are interested in an analog of the electric-magnetic duality.  
While we have a number of exact electrically charged solutions in 
(2+1)-dimensional gravity theories, for example, as can be found in 
Ref. \cite{cm}, magnetically charged solutions are relatively less 
understood.  If we find an analog of the electric-magnetic duality, we 
can use it to find magnetically charged solutions from the known 
electrically charged solutions.  In case of the BTZ theory, the 
magnetically charged solutions were obtained in \cite{welch}.  As 
their results show, the properties of the magnetic solutions
are too different from the electric solutions to immediately uncover 
any relations between them.  However, the possibility of an analog of 
the electric-magnetic duality was suggested in further studies, for 
example, in \cite{well} for the BTZ theory and in \cite{emd3} for the 
case of the Einstein-Maxwell-Dilaton theory without the cosmological 
constant term.      

In this note, we find an analog of the electric-magnetic duality for 
the static and rotationally symmetric solutions of the theories given by 
the (2+1)-dimensional action
\begin{equation}
I = \int d^3 x \sqrt{g^{(3)}} ( R^{(3)} - 
\frac{1}{2} g^{ (3) \alpha \beta} \partial_{\alpha} f \partial_{\beta} f
 + \Lambda e^{b f} + \frac{1}{4} e^{ \chi f} F^2 )
\label{start}
\end{equation}
where $R^{(3)}$, $f$, $F$ denote the (2+1)-dimensional scalar curvature, 
the dilaton field and the curvature two form for a $U(1)$ gauge field,
respectively.  We use $(+--)$ signature for the (2+1)-dimensional
metric $g^{(3)}_{\alpha \beta}$.
We also have the cosmological constant $\Lambda$ and two 
parameters $b$ and $\chi$, the specification of which gives us a 
particular gravity theory.  For example, the choice $b = \chi = 0$ 
produces the BTZ theory, while the choice $b = -\chi = \sqrt{2}$ 
yields the (2+1)-dimensional low energy string effective 
action after a rescaling of the metric $g^{(3)}_{\alpha \beta}$.  
Under the assumption of the rotational symmetry, we can write the
(2+1)-dimensional metric as
\begin{equation}
ds^2 = g_{\alpha \beta} dx^{\alpha} dx^{\beta} - 
       e^{-4 \phi} d \theta^2 ,
\label{metric}
\end{equation}
where the two-dimensional longitudinal metric $g_{\alpha \beta}$
and the conformal factor $\phi$ of the angular part of the metric 
is independent of the azimuthal angle $\theta$.  
We choose to describe the resulting 2-dimensional longitudinal
geometry of the space-time in terms of a conformal gauge, thereby 
setting $g_{\alpha \beta} dx^{\alpha} dx^{\beta} = - \exp ( 2 \rho )
dx^+ dx^-$.  We can now reduce the action Eq. (\ref{start}) 
to a class of 2-dimensional dilaton gravity 
theories \cite{banks} by integrating out the $\theta$ 
coordinate \cite{pk}.
In this process, $(\pm , \theta )$ components of the Einstein
equations, that can not be captured in the resulting
2-dimensional action, reduces to a condition
\begin{equation}
F_{+-} F_{\theta \pm} = 0 . 
\label{cond}
\end{equation}
To get static solutions, we assume all the physical
variables in our consideration depends only on a space-like 
variable $x = x^+ + x^-$.  This in particular implies
$F_{+ \theta} = F_{- \theta}$ or, in other words, the two
form curvature $F$ of the gauge field consists of the 
purely electric field $F_{-+}$ and the purely magnetic field 
$F_{\pm \theta}$.  Eq. (\ref{cond}) therefore shows the solutions 
of our problem have either electric charge or magnetic charge, 
but not both at the same time.  For our further consideration,
we introduce a field $A$ that, in electrically charged case,
is defined to satisfy $F_{-+} = dA/dx$, and 
$F_{\pm \theta} = dA/dx$ in magnetically charged case.
Explicitly, the action for the electrically charged sector
can be written as
\begin{equation}
I_e = - \int d \bar{x} e^{\rho - 2 \phi + \frac{b}{2} f}
\left[ 2 \phi^{\prime} \rho^{\prime} + \frac{1}{4} f^{\prime 2}
- \frac{1}{4} e^{\chi f - 2 \rho} A^{\prime 2} 
+ \frac{\Lambda}{4} \right] ,
\label{ele}
\end{equation}
while we have the following action for the magnetically charged
sector.
\begin{equation}
I_m = - \int d \bar{x} e^{\rho - 2 \phi + \frac{b}{2} f}
\left[ 2 \phi^{\prime} \rho^{\prime} + \frac{1}{4} f^{\prime 2}
+ \frac{1}{4} e^{\chi f  + 4 \phi} A^{\prime 2} + 
\frac{\Lambda}{4} \right] .
\label{mag}
\end{equation}
Here we introduce a new spatial coordinate $\bar{x}$ via
$d\bar{x}= \exp ( \rho +bf/2 )dx $ and the prime denotes the
differentiation with respect $\bar{x}$.  We should also impose
the static version of the gauge constraints for each case that results 
from our choice of a conformal gauge.

The underlying symmetry of the theories in our consideration
is most apparent when we introduce a set of field redefinitions
to new fields $X$, $Y$ and $u$ given by 
\begin{equation}
\left( \begin{array}{c}  X \\ Y \\ u \end{array} \right) 
= T \left( \begin{array}{c} \rho \\ \phi \\ f \end{array} \right)
\label{fred}
\end{equation}
where the matrix $T$ is
\[ T  = 
\left( \begin{array}{ccc} 
     \frac{b}{\sqrt{2}} &  0     & \frac{1}{\sqrt{2}}  \\
      \epsilon \frac{2-b^2}{\sqrt{2 | 4-b^2 |} } 
    & \epsilon \frac{2 \sqrt{2}}{\sqrt{|4-b^2 |}}
    & - \epsilon \frac{b}{\sqrt{2 | 4-b^2 | } }  \\
      1          &  -2   &  \frac{b}{2} \end{array} \right) . \]
We define $\epsilon \equiv (4- b^2 )/ |4-b^2 |$, which becomes $+1$ for 
$ | b| < 2$ and $-1$ for $ |b | > 2$.  The determinant of the
matrix $T$ is $- | b^2 - 4|^{1/2}$.  Therefore as long as the
parameter $b^2$ is not $4$, the field redefinitions, Eq. (\ref{fred}),
are well-defined.  In what follows, we find it convenient to use a 
vector notation $\vec{X} = (X, Y)$ 
on the 2-dimensional space of fields $X$ and $Y$ with an inner product 
$\vec{X} \cdot \vec{X} = X^2 + \epsilon Y^2$.  The action for both the 
electric sector and the magnetic sector, then, becomes 
\begin{equation}
I=-\int d\bar{x} e^u (  \frac{1}{2}
     \vec{X}^{\prime} \cdot \vec{X}^{\prime} 
     -\frac{u^{\prime 2}}{4-b^2}
      +   \frac{p}{4} 
 \exp ( \vec{d} \cdot \vec{X} -\frac{2\chi b+4}{4-b^2}u ) 
  A^{\prime 2}  + \frac{\Lambda}{4} ) ,
\label{ie}      
\end{equation}
while the gauge constraints can be written as
\begin{equation}
\frac{ \vec{X}^{\prime} \cdot \vec{X}^{\prime}}{2} 
-\frac{ u^{\prime 2}}{4-b^2}
+  \frac{p}{4}\exp 
( \vec{d} \cdot \vec{X} - \frac{2\chi b+4}{4-b^2}u ) 
A^{\prime 2} -\frac{\Lambda}{4} = 0.
\label{econ}
\end{equation}
upon using the static equations of motion.  The only difference
between the action for the electrically charged sector
and the magnetically charged sector is the choice of a 
2-dimensional vector $\vec{d}$ and a parameter $p = \pm 1$.
For the electrically
charged sector, we have the vector $\vec{d} = \vec{d}_e$ where
\begin{equation}
\vec{d}_e = \sqrt{2} ( \chi , - \epsilon \frac{2 + \chi b }
{\sqrt{|4-b^2 | } } )  
\label{evec}
\end{equation}
along with $p = -1$.  For the magnetically charged sector,
we have $\vec{d} = \vec{d}_m$ where
\begin{equation}
\vec{d}_m = \sqrt{2} ( b + \chi ,  \epsilon 
\frac{2 - b^2 - \chi  b }{\sqrt{|4-b^2 | } } )
\label{mvec}
\end{equation}
along with $p = +1$.
We notice that the action (\ref{ie}) has a symmetry under the
transformation $\bar{x} \rightarrow \bar{x} + \alpha$ for an
arbitrary constant $\alpha$.  The role of the gauge constraint
(\ref{econ}) is to set the Noether charge of this symmetry to
zero \cite{pk}. 

Eq. (\ref{ie}) manifestly shows the 
$O(2)$ symmetry for $|b| < 2$ ($\vec{X} \cdot \vec{X}
= X^2 + Y^2 $) and $O(1,1)$ symmetry
for $|b| > 2$ ($\vec{X} \cdot \vec{X} = X^2 - Y^2 $)
when we set $A = 0$, i.e., when there is no
$U(1)$ charge.  (When $\Lambda = 0$, we have an enhanced symmetry.
Then, by a redefinition of $\bar{x}$ and a rescaling of $u$, 
$O(2,1)$ symmetry becomes manifest.  We also note the $O(2)$ symmetry 
was first observed in \cite{caldoni}.) 
These symmetries correspond to the rotation
in $(X, Y)$ space.  Furthermore, together with the two translational
symmetries of $X$ and $Y$, which are also present when $A=0$,
they constitute the 2-dimensional Euclidean Poincare group 
or the 2-dimensional Minkowskian Poincare group, depending on
the value of $b$.  When we introduce an electric charge, we end up 
choosing a particular vector $\vec{d}_e$ in the 2-dimensional 
$(X, Y)$ space.  This  breaks the Poincare invariance to the $Z_2$ 
symmetry, that corresponds to the reflection about the electric axis 
$\vec{d}_e$, and a single translational 
symmetry of the fields $(X,Y)$ along 
the direction perpendicular to the vector $\vec{d}_e$.
A similar story holds when we introduce a magnetic charge;
we break the uncharged Poincare invariance to the $Z_2$ symmetry
about the magnetic axis $\vec{d}_m$ plus one translational
symmetry.  

We concentrate on the case $0 \le b < 2$, which contains two prime
examples of our concern, namely, the BTZ model ($b = \chi = 0$)
and the target space effective action of the string theory
($b = -\chi = \sqrt{2}$).  For uncharged cases, we then have
the $O(2)$ symmetry.  A key observation that shows the existence
of an analog of the electric-magnetic duality is that the equality
\begin{equation}
 \vec{d}_e \cdot \vec{d}_e = \vec{d}_m \cdot \vec{d}_m = 
 8 \frac{1 + b \chi + \chi^2}{4 - b^2} 
\end{equation}
holds.  Therefore, by the reflection about the axis that bisects
the electric axis and the magnetic axis, we can transform the
electric action exactly into the magnetic action
and vice versa.  Of course, just as in the
case of the electric-magnetic duality of the 4-dimensional
Minkowskian space-time, we need to flip the sign of the
$A^{\prime 2}$ term along with the reflection.  This
$Z_2$ transformation, which is also a subgroup of the uncharged
$O(2)$ group, is an analog of the electric-magnetic duality
in case of the (2+1)-dimensional gravity theories.

An important application of our results is to obtain 
magnetically charged solutions from the known electrically
charged solutions.  For this purpose, it is convenient to work out 
the representation of the above transformations that acts on the
space of $(\rho , \phi, f)$ fields.  Using the matrix $T$ and the
above considerations, we find
\begin{eqnarray}
 \left( \begin{array}{c} 
  \rho_m \\ \phi_m \\ f_m  \end{array} \right)
 = \tau_{me} \left( \begin{array}{c}
  \rho_e \\ \phi_e \\ f_e  \end{array} \right) ,  ~~~~ &&
 \left( \begin{array}{c} 
  \rho_e \\ \phi_e \\ f_e  \end{array} \right)
 = \tau_{em} \left( \begin{array}{c}
  \rho_e \\ \phi_e \\ f_e  \end{array} \right)    \nonumber \\
  \left( \begin{array}{c} 
  \rho_e \\ \phi_e \\ f_e  \end{array} \right)
 = \tau_{ee} \left( \begin{array}{c}
  \rho_e \\ \phi_e \\ f_e  \end{array} \right) ,  ~~~~ &&
 \left( \begin{array}{c} 
  \rho_m \\ \phi_m \\ f_m  \end{array} \right)
 = \tau_{mm} \left( \begin{array}{c}
  \rho_m \\ \phi_m \\ f_m  \end{array} \right)
\label{expt}
\end{eqnarray}
where the matrices are given by
\[
\tau_{me} = \frac{1}{1 + b \chi + \chi^2}
\left( \begin{array}{ccc}
  (b+ \chi )^2     &  -2     &   b + \chi \\
  - \frac{1}{2}    & \chi^2  & \frac{\chi}{2} \\
  -2(b + \chi )    & -4 \chi & -1 + b \chi + \chi^2 \end{array} \right)
\]
\[
\tau_{em} = \frac{1}{1 + b \chi + \chi^2}
\left( \begin{array}{ccc}
  \chi^2      &  -2          &  -\chi  \\
 -\frac{1}{2} & (b+\chi )^2  & - \frac{b+ \chi}{2} \\
  2 \chi      & 4(b+ \chi )  & -1 + b \chi + \chi^2 \end{array} \right)
\]
\[
\tau_{ee} = \frac{1}{1 + b \chi + \chi^2}
\left( \begin{array}{ccc}
               1    &  -2(b+\chi )^2    &   b + \chi \\
 -\frac{\chi^2}{2}  &      1            & \frac{\chi}{2} \\
           2\chi    & 4(b + \chi )      & -1 + b \chi + \chi^2 \end{array}
\right)
\]
\[
\tau_{mm} = \frac{1}{1 + b \chi + \chi^2}
\left( \begin{array}{ccc}
               1     &  -2\chi^2  &  - \chi \\
 -\frac{(b+\chi )^2}{2} &      1     & -\frac{b+\chi}{2} \\
      -2(b+\chi )    & -4 \chi    & -1 + b \chi + \chi^2 \end{array}
\right)
\]
The subscripts $e$ and $m$ for each field represent the solutions 
in the electrically charged sector and in the magnetically charged 
sector, respectively.  We can straightforwardly verify that 
$\tau_{em}$ transforms Eq. (\ref{mag}) into Eq. (\ref{ele}), 
$\tau_{me}$ transforms Eq. (\ref{ele}) into Eq. (\ref{mag}),
$\tau_{ee}$ leaves Eq. (\ref{ele}) invariant, and 
$\tau_{mm}$ leaves Eq. (\ref{mag}) invariant.
Since $\tau_{ee}$ ($\tau_{mm}$) denotes
the self-duality within the electrically charged sector (magnetically
charged sector), they satisfy $\tau_{ee}^2 = \tau_{mm}^2 = 1$
and $\det \tau_{ee} = \det \tau_{mm} = -1$.  We can also 
straightforwardly verify
that $\tau_{me} \tau_{em} = \tau_{em} \tau_{me} = 1$ and
$\det \tau_{me} = \det \tau_{em} = 1$.  Another interesting
observation is the simultaneous transformation $b + \chi \rightarrow
- \chi$ and $\chi \rightarrow - ( b+ \chi )$ exchanges $m$ and $e$
subscripts.  Using Eq.(\ref{expt}), it is straightforward to
obtain magnetically charged solutions from the known electrically
charged solutions.  From \cite{cm}, some exact electrically
charged solutions for the $b = - \chi$ case are available.  We 
recast them in the conformal gauge, apply $\tau_{me}$, and 
go back to the original gauge, to obtain the following exact
magnetically charged solutions.  The metric is computed to be
\begin{eqnarray}
ds^2 =  (\frac{r^2}{l^2} - M)  dt^2
&& - r^2 ( \frac{r^2}{l^2} - M)^{b^2 /2 -1 }
\left[ r^2 + N ( \frac{r^2}{l^2} - M )^{b^2 /4 } - N \right]^{-1} 
dr^2 \nonumber \\
&& -  \left[r^2 + N ( \frac{r^2}{l^2} - M )^{b^2 /4} - 
N \right] d \theta^2
\label{me}
\end{eqnarray}
and the dilaton field $f$ turns out to be
\begin{equation}
 e^{b f} = k^{b^2 /2} e^{-b f_1}  ( \frac{r^2}{l^2} - M )^{-b^2 /2 }
\label{dila}
\end{equation}
where $l$, $M$, $k$, $N$ and $f_1$ are constants.  In case of the BTZ
theory, $b = \chi = 0$, we take the limit $b \rightarrow 0$ of 
Eqs. (\ref{me}) and (\ref{dila}) setting
$M \rightarrow M_0$, $l \rightarrow l_0$
and $N \rightarrow 4 Q_M^2 / b^2 $.   The dilaton field in this
case becomes a constant $f = - f_1$ and the metric becomes
\[ ds^2 = (\frac{r^2}{l_0^2} - M_0)  dt^2
- \frac{r^2}{ ( \frac{r^2}{l_0^2} - M_0)
( r^2 + Q_M^2 \ln | \frac{r^2}{l_0^2} - M_0 |) }dr^2
-  ( r^2 + Q_M^2 \ln |\frac{r^2}{l_0^2} - M_0  |  )d \theta^2 , 
\]
which is exactly the same as the solutions of \cite{welch}.  
For other cases
including the most important string effective theory,
$ b = -\chi = \sqrt{2}$, we obtain new non-trivial magnetically 
charged solutions.  It is interesting that the drastically different
solutions in each sector are related by an analog of the usual
electric-magnetic duality unlike the (3+1)-dimensional case.  
In fact, when $\Lambda = 0$, it was found in \cite{emd3} that the
magnetically charged solutions in \cite{barrow} is related
to the electrically charged solutions in \cite{gott} via an
analog of the electric-magnetic duality.  Actually, we can also 
generate, for example, the dual of the known electrically charged 
solutions via $\tau_{ee}$.  We plan to address the detailed study of 
the new solutions and the relationships between solutions related 
by the dual transformations in a future publication.

We can understand more of the structure of the transformations we find
so far by noting the relation
\begin{equation}
\cos \alpha = \frac{\vec{d}_m \cdot \vec{d}_e}{\vec{d}_m \cdot \vec{d}_m}
= \frac{ \chi^2 + b \chi  + b^2 /2 -1 }{1 + b \chi + \chi^2 }
\label{angle}
\end{equation}
that shows the angle $\alpha $ between the electric and the magnetic 
vector for a given set of parameters $(b , \chi)$.  By varying the 
value of $\alpha$ from zero to $\pi$, we form a disjoint collection
of paths in $(b, \chi )$ space.  One distinctive case is when a
path is represented by $b + 2 \chi = 0$.  We then have
$\cos \alpha = -1$, which means the $\vec{d}_e$ points to the 
opposite direction to the vector $\vec{d}_m$.  The BTZ theory
with $b = \chi = 0$ belongs to this case.  Clearly, the $\tau_{me}$
matrix satisfy $\tau_{me}^2 = 1$, thereby $\tau_{me} = \tau_{em}$,
since the two rotations, each of them by 180 degrees, 
combine to become the 
identity operation.  We also have $\tau_{mm}= \tau_{ee}$, since they 
represent the reflection about the electric and magnetic axis,
each of which points to the opposite direction to each other.
The matrices $\tau_{me}$ and $\tau_{ee}$, that commute with
each other in this case, generate the direct sum of two $Z_2$ groups,
the Klein four group.  In general, when $\alpha = 2 \pi / n$ 
for an integer $n \ge 2$, the smallest positive integer $m$ 
that satisfies
$\tau_{me}^m = 1$ is $n$.  Thus, together with $\tau_{ee}$ and
$\tau_{mm}$, $\tau_{me}$ generates the $D_{n}^*$, the symmetry
group of $n$-gon.  The most interesting case is when $n=4$ that
gives a situation where the vector $\vec{d}_e$ is perpendicular
to the vector $\vec{d}_m$.
In this case, $\alpha = \pi /2$, thus
\begin{equation}
\chi^2 + b \chi + b^2 /2 -1 = 0
\label{some}
\end{equation}
and the 8-element nonabelian group $D_4^*$ is isomorphic to $O(2, Z)$.  
The low energy string effective theory with $b = - \chi = \sqrt{2}$
satisfies Eq.(\ref{some}).  Since $O(2,Z)$ is both a natural subgroup
of $O(2,2,Z)$ and $O(2)$, the transformations we find may be 
considered as a natural subgroup of the $T$-duality from 
the string theory.
For other values of $n$, including $n=2$ for the BTZ theory, 
$\tau_{em}$, $\tau_{ee}$ and $\tau_{mm}$ generate a group different
from $O(2,Z)$.  When combined with $b = -\chi$, which is clear from the
fact that the cosmological constant term results from the
closed string sector and the $U(1)$ field, from the tree
level open string sector, Eq.(\ref{some}) uniquely determines
$b = -\chi = \pm \sqrt{2}$.  Both sets of the parameters correctly
give the kinetic term of the dilaton field $f$ for the low energy
effective string theory after a transformation $f \rightarrow \pm f$
in Eq.(\ref{start}).  This shows an interesting possibility; the 
parameters (for example, $b$ and $\chi$ in our case) appearing in the 
low energy string effective action may be determined to some extent by 
the requirement of preservation of the $T$-duality.  A 
similar analysis in higher dimensional cases to the 
one given here may further verify this idea and the work in 
this regard is in progress.

%\begin{acknowledgments}
%\end{acknowledgments}

\end{document}